\documentclass[aps,prl,reprint, amsmath, amssymb,superscriptaddress,nofootinbib]{revtex4-1}
\usepackage{bm}
\usepackage[utf8]{inputenc}
\usepackage[T1]{fontenc}
\usepackage[retainorgcmds]{IEEEtrantools}
\usepackage{graphicx}
\usepackage{mathrsfs}
\usepackage{amsmath}
\usepackage{indentfirst}
\usepackage{amssymb}
\usepackage{color}
\usepackage{amsfonts}
\usepackage[framemethod=TikZ]{mdframed}
\usepackage{dsfont}
\usepackage{times,txfonts}
\usepackage{appendix}
\usepackage{nicefrac}
\usepackage{enumitem} 
\usepackage{braket}
\usepackage{mathtools}
\usepackage[colorlinks=true,linkcolor=blue,urlcolor=blue,citecolor=blue,pdfusetitle]{hyperref}
\usepackage{soul}

\mdfdefinestyle{MyFrame}{%
    linecolor=black,
    outerlinewidth=1pt,
    roundcorner=0pt,
    innertopmargin=0.77\baselineskip,
    innerbottommargin=0.75\baselineskip,
    innerrightmargin=14pt,
    innerleftmargin=14pt,
    backgroundcolor=gray!10!white}

\newcommand{\dbar}{{\mathchar'26\mkern-11mud}}

\newlength{\eqboxstorage}

\begin{document}

\title{Coherent energy fluctuation theorems: theory and experiment}

\author{K.  Khan}
\affiliation{Federal University of Rio de Janeiro, Rio de Janeiro, RJ, Brazil}

\author{J. Sales Ara\'ujo}
\affiliation{Federal University of Rio de Janeiro, Rio de Janeiro, RJ, Brazil}

\author{W. F. Magalh\~aes}
\affiliation{Departamento de F\'isica, Universidade Federal da Para\'iba, 58051-900 Jo\~ao Pessoa, PB, Brazil}

\author{G. H. Aguilar}
\email{gabo@if.ufrj.br}
\affiliation{Federal University of Rio de Janeiro, Rio de Janeiro, RJ, Brazil}

\author{B. de Lima Bernardo}
\email{bertulio.fisica@gmail.com}
\affiliation{Departamento de F\'isica, Universidade Federal da Para\'iba, 58051-900 Jo\~ao Pessoa, PB, Brazil}
\begin{abstract}

Heat, work and entropy production: the statistical distribution of such quantities  are constrained by the fluctuation theorems (FT), which reveal crucial properties about the nature of non-equilibrium dynamics. In this paper we report theoretical and experimental results regarding two FT for a new quantity, named coherent energy, which is an energy form directly associated with the coherences of the quantum state. We also demonstrate that this quantity behaves as a thermodynamic arrow of time for unitary evolutions, that is, in the absence of entropy production.  The experiment is implemented in an all-optical setup in which the system is encoded in the polarization of one photon of a pair. The FT are demonstrated using the two-point measurement protocol,  executed using the other photon of the pair,   allowing to  assess the probability distributions directly from the outcomes of the experiment. 

\end{abstract}

\maketitle 

%
%
{\it Introduction:}
%
%
Theoretical descriptions of irreversible processes are notoriously challenging, especially when the system is driven far from equilibrium. In this regime, the so called fluctuation theorems, which can be seen as generalizations of the second law of thermodynamics, provide a unique insight into the nature of the macroscopic irreversibility from time-symmetric fundamental laws \cite{bustamante,jar}. Among the many types of FT discussed in the literature \cite{evans,gallavotti,crooks,jarzynski,hatano,seifert,kawai,esposito2}, the work FT are of special importance both from a historical and a systematic viewpoint. These theorems relate the work $W$ performed in nonequilibrium processes with free-energy differences $\Delta F$, which is a feature of reversible transformations. One of these FT is the Crooks theorem \cite{crooks}, which states that the probability  $P(W)$ of realizing  a work $W$ on a system (initially at equilibrium with an inverse temperature $\beta$) in a given nonequilibrium process is related to the probability  $\Tilde{P}(-W)$ of realizing a work $-W$ with the time-reversed process (again with the system initially at equilibrium) according to the relation $P(W)/\Tilde{P}(-W) = \exp[\beta(W-\Delta F)]$. Another example is the famous Jarzynski equality \cite{jarzynski}, $\langle \exp [- \beta W] \rangle = \exp [- \beta \Delta F]$, which can be derived by simple integration of Crooks theorem. These FT have been experimentally confirmed by using different classical small systems such as colloidal particles \cite{wang}, biopolymers \cite{liphardt,collin}, mechanical oscillators \cite{douarche}, and Brownian particles \cite{toyabe,martinez}.  

FT have also been studied in the presence of quantum effects, such as intrinsic statistical uncertainty of quantum observables \cite{esposito,campisi}. These quantum fluctuation theorems (QFT) have been derived in a similar fashion as for the classical case \cite{kurchan,tasaki,talkner,deffner} and experimentally verified in nuclear-magnetic-ressonace \cite{batalhao} and trapped ions \cite{an},  nitrogen-vacancy centers \cite{gomez}, and a quantum computer \cite{solfaneli}. In such cases, the notion of trajectory is not accessible without directly interfering in the system, and the definition of work ends up being nontrivial. This problem is usually addressed in the framework of the ``two-point measurement'' (TPM) protocol \cite{esposito,campisi}. Here,  the work is associated with the difference between the outcomes of two energy projective measurements, one before and the other after the evolution of the closed, driven system.  The TPM scheme has also been used as a test bed in theoretical investigations of the influence of coherence in the energetics of non-equilibrium transformations \cite{Mingo19, Kwon19, Lostaglio18, Gherardini21}.
For instance, a QFT for entropy production has been deduced by performing  a modified TPM protocol, where the projective  measurements are performed in the eigenbasis of the density matrix of the system \cite{santos}. Conversely, it has been shown that the irreversible work in a quantum process can be split into a coherent and a incoherent contributions both fulfilling a QFT separately \cite{francica}. However, despite such efforts devoted to understand this scenario, theoretical insights and experimental demonstrations of QFT emphasizing the contribution of coherence are still missing.


 In this paper, we investigate theoretically and experimentally two QFT for the coherent energy, a quantity defined as the energy transferred to or from a quantum system, which is  directly linked to a change in coherence \cite{bernardo,bernardo2}. The proposed QFT  provide key information about the fluctuations of the coherent energy in nonequilibrium transformations, and hence in the dynamics of coherence. More specifically, we show that the mean coherent energy determines the ``arrow of time" for unitary transformations acting on thermal states, i.e.,  when the entropy production is null \cite{deffner,seifert}. 
  Our all-optical experiment implements the TPM protocol, nevertheless, instead of making measurements in sequence on the same system, the two outcomes are obtained through separate measurements  onto a pair of entangled photons \cite{aguilar}. This experimental method circumvents the destructive nature of photo-detections at the first measurement of the TPM  and allows  direct assess to the probability distribution of coherent energy.


{\it Coherent energy:}  Let us consider an arbitrary quantum system, whose Hamiltonian is given by $\hat{H} = \sum_{n} E_{n} \ket{n}\bra{n}$, where the $n$-th energy eigenvalue and eigenstate are $E_{n} = \braket{n|\hat{H}|n}$ and $\ket{n}$. The density operator of the system is $\hat{\rho} = \sum_{k} \rho_{k} \ket{k}\bra{k}$, with $\rho_{k} = \braket{k|\hat{\rho}|k}$ and $\ket{k}$ being the eigenvalues and eigenkets. In this framework, it was demonstrated that the internal energy of the system, defined as the average of the Hamiltonian, $U = \langle \hat{H} \rangle =$ tr$\{\hat{\rho} \hat{H} \}$, could change through the transfer of three types of energy: {\it work}, {\it heat} and {\it coherent energy} \cite{bernardo}. As in the classical case \cite{kittel, callen, landau}, the work $W$ is considered as ``the variation in the internal energy that results from changes in the generalized coordinates of the system'', and the heat $Q$ is understood as ``the variation in the internal energy that accompanies entropy change''. These two quantities define the first law of classical thermodynamics simply as $dU = \dbar W + \dbar Q $. By contrast, the coherent energy $\mathcal{C}$ was introduced in the quantum thermodynamic context as ``the change in the internal energy accompanied by coherence change''. In fact,  in addition to work and heat exchanges, the amount of coherent energy also needs to be taken into account in the study of quantum processes \cite{bernardo2}. Then, it was proposed that the first law of thermodynamics should be redefined as $dU = \dbar W + \dbar Q + \dbar \mathcal{C}$, in order to encompass the role played by coherence \cite{bernardo}.

The above considerations allows us to quantitatively describe the behavior of work, heat and coherent energy for arbitrary finite-time quantum transformations respectively as \cite{bernardo}:  
\begin{equation}
\label{1}
W(t)  =  \sum_{n} \sum_{k} \int_{0}^{t}  \rho_{k} |c_{n,k}|^{2} \frac{d E_{n}}{dt'} dt',
\end{equation}
\begin{equation}
\label{2}
Q(t)  =  \sum_{n} \sum_{k} \int_{0}^{t}  E_{n} |c_{n,k}|^{2} \frac{d \rho_{k}}{dt'} dt',
\end{equation}    
\begin{align}
\label{3}
\mathcal{C}(t)
&= \sum_{n}\sum_{k} \int_{0}^{t} (E_{n} \rho_{k}) \frac{d}{dt'} |c_{n,k}|^{2} dt',
\end{align}
where $c_{n,k} = \bra{n}k\rangle$. In this scenario, we see that realization of work takes place in processes in which there is a change in the system's spectrum $\{E_{n}\}$, and heat exchange occurs when there is some modification in the eigenvalues of the density matrix, $\{ \rho_{k} \}$. In turn, a transfer of coherent energy only exists when the coefficients $\{ |c_{n,k}| \}$ are time-dependent, i.e., when the quantum transformation involves a change in the quantum coherence of the system. The internal energy change in this perspective is given by $\Delta U(t) = W(t) + Q(t) + \mathcal{C}(t)$, keeping in mind that all these functions represent the time evolution of the averages of each type of energy, when many runs of the same experiment are realized.

{\it Fluctuation theorems for coherent energy:} We now discuss the probabilistic properties of the coherent energy, which will lead us to two QFT. To this end, we study the coherent energy transfer in the usual TPM scheme \cite{esposito,campisi}. In this approach, the coherent energy will correspond to the difference between the outcomes of two projective energy measurements made before and after a quantum process in which only energy accompanied by coherence change is transferred to or from the system. In such a case, the quantum process must satisfy two conditions: i) it must be unitary in order to guarantee that the entropy change of the system, and hence the heat exchange, is zero, and ii) the energy levels of the system must be unchanged to guarantee that no work is realized. This process will be represented by an unitary operator $\hat{\mathcal{U}}(0,\tau)$, which acts from an initial time $t =0$ until a final time $t =\tau$. 

We assume that at $t=0$ the system is in thermal contact with a heat bath at inverse temperature $\beta$, so that it is in the thermal state $\hat{\rho}(0) = e^{-\beta E_{n}}/Z \ket{n}\bra{n}$, where $Z=$ tr$(e^{-\beta \hat{H}})$ is the partition function. Basically, this is the time when the first energy measurement of the TPM protocol is performed, such that each possible outcome $E_{n}$ can be obtained with probability $P_{n} = e^{-\beta E_{n}}/Z$. After this first measurement, the system evolves unitarily under the action of $\hat{\mathcal{U}}(0,\tau)$ before being submitted to the second energy measurement at $t=\tau$. By denoting the first and second measurement outcomes, respectively, by $E_{n}$ and $E_{m}$, the coherent energy transferred in each process is simply given by $\mathcal{C} = E_{m} - E_{n}$
.

\begin{figure}
    \includegraphics[width=8.5cm]{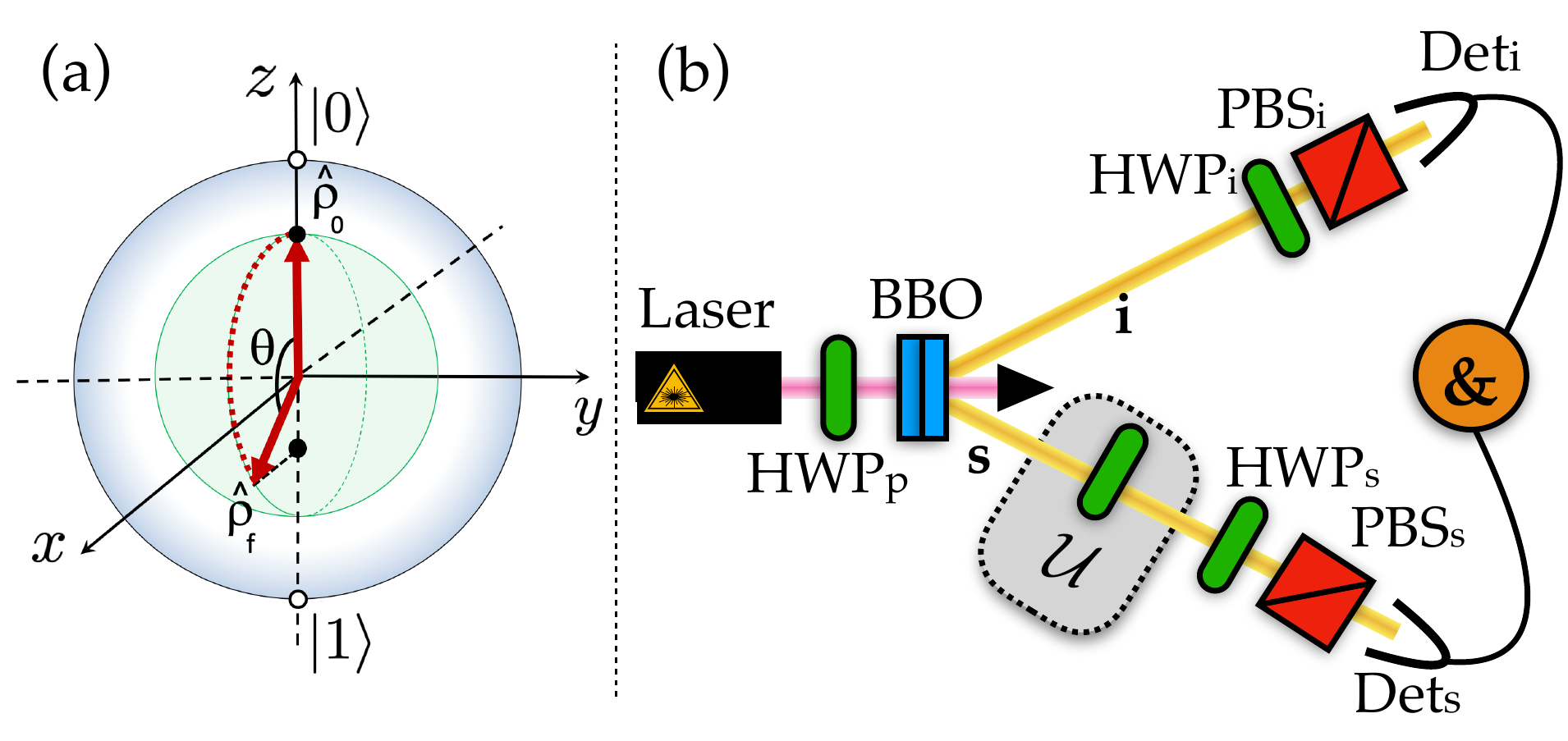}
    \caption{(a) Bloch sphere representation of a qubit thermal state $\hat{\rho}_{0}$ undergoing an unitary operation. The size of the vector is kept fixed during the transformation but its z-component diminishes independently of $\mathcal{U}$. This indicates the arrow of time.  (b) Experimental setup:  Entangled photon pairs are created in a state $\ket{\psi_0}$ in the modes s and i. Photons in s pass through a HWP, which apply an unitary operation $\mathcal{U}$. Both photons are detected after projective measurement in their polarization. }
    \label{fig:setup}
\end{figure}

Let us now study the probability
of obtaining a particular value for the coherent energy $\mathcal{C}$ in a single realization of the TPM protocol. Of course, this is the joint probability of measuring $E_{n}$ at $t=0$ and $E_{m}$ at $t=\tau$. Thus, the coherent energy probability distribution is given by
\begin{align}
\label{4}
P(\mathcal{C})
= \sum_{n,m} P_{n} |\bra{m}\hat{\mathcal{U}}(0,\tau)\ket{n}|^2 \delta[\mathcal{C} - (E_{m} - E_{n})],
\end{align}
where $\delta(x)$ is the Dirac delta function. Due to the large number of possible transitions $\ket{n} \rightarrow \ket{m}$, it is convenient to work with the characteristic function, which is the Fourier transform of $P(\mathcal{C})$,  
\begin{align}
\label{5}
\chi (q) = \langle e^{i q \mathcal{C}} \rangle = \int_{- \infty}^{\infty} P(\mathcal{C}) e^{i q \mathcal{C} } d \mathcal{C}.
\end{align}
The substitution of Eq.~(\ref{4}) into Eq.~(\ref{5}) leads to the simple relation
$\chi (q) = tr \{\hat{\mathcal{U}}^{\dagger} e^{i q \hat{H}} \hat{\mathcal{U}} e^{-i q \hat{H}} \hat{\rho}(0) \}$, where we omitted the time dependence of the unitary operators. At the same time, we notice that $\chi (q = i \beta) = \langle e^{ - \beta \mathcal{C}} \rangle$, which provides our first QFT,
\begin{align}
\label{6}
\langle e^{ - \beta \mathcal{C}} \rangle = 1.
\end{align}
This integral fluctuation theorem (IFT)  reveals a unique feature of the dynamics of quantum coherence in non equilibrium thermal processes. From this result, together with Jensen's inequality, we also verify that $\langle e^{ - \beta \mathcal{C}} \rangle \geq e^{- \beta { \langle  \mathcal{C} \rangle}} $, and hence that $\langle \mathcal{C} \rangle \geq 0$, which means that, in the described process, the coherent energy never decreases on average. In other words, this result implies that, on average, it is not possible to reduce the internal energy of a thermal quantum system that undergoes an arbitrary unitary operation in which the energy levels are kept fixed. This is analogous to what happens with the mean entropy production in more general transformations. In this respect, the arrow of time is indicated by the non-negativity of the mean entropy production. However, it is worth mentioning that for unitary transformations, such as the studied here, the entropy production is null and the definition of the time arrow is still missing. In the Supplementar Material \cite{Supplementar}, we prove  that coherent energy is an excellent tool for this definition. To explain this pictorially for the simple case  of two-level systems, we use the  Bloch sphere  representation in  Fig. \ref{fig:setup}(a). Here,  a  thermal state $\hat{\rho}_0$ is represented by a vector pointing upwards, whose size is kept fixed  after an unitary transformation $\mathcal{U}$ (rotation operation).  For this case, the mean coherent energy $ \langle \mathcal{C} \rangle $ is directly linked to the reduction of the $z$-component of the vector. One can see that for  any $\mathcal{U}$, this component can only be reduced, indicating the non-negativity of $\langle \mathcal{C} \rangle$. This univocally defines the time-evolution direction,  establishing a direct connection between the coherent energy and the arrow of time for unitary transformations.  

Conversely, another fundamental ingredient in the  context of QFT is the principle of time reversibility of closed quantum systems, $\hat{\mathcal{U}}(0,\tau) = \hat{\mathcal{U}}^{\dagger}(\tau,0) = \hat{\mathcal{U}}^{-1}(\tau,0)$ \cite{sakurai}, which allows the derivation of a second QFT. Indeed, by using this fact and that $E_{n} = E_{m} - \mathcal{C}$ in Eq.~(\ref{4}), we can write 
\begin{align}
\label{7}
P(\mathcal{C})
&=e^{ \beta \mathcal{C}} \sum_{n,m} P_{m} |\braket{n|\hat{\mathcal{U}}^{-1}(\tau,0)|m}|^2 \delta[- \mathcal{C} - (E_{n} - E_{m})],
\end{align}
which can be rewritten as
\begin{align}
\label{8}
\frac{P(\mathcal{C})}{\Tilde{P}(-\mathcal{C})} = e^{\beta \mathcal{C}}.
\end{align}
Here, $\Tilde{P}(-\mathcal{C})$ is the probability distribution of obtaining a coherent energy $-\mathcal{C}$ in a TPM protocol which is the backward (time-reversed) protocol of that described in Eq.~(\ref{4}). The result of Eq.~(\ref{8}) is a  detailed fluctuation theorem (DFT) for the coherent energy, which has an important physical meaning. It says that the probability of a thermal equilibrium system absorbing a given amount of coherent energy $\mathcal{C}$, in a process where no work is realized and no heat is exchanged, is larger by a factor $e^{ \beta \mathcal{C}}$ than the probability of releasing the same amount in the time-reversed process.



%
%

%
{\it Experimental setup:} The experimental setup to investigate the FT for coherent energy  is shown in Fig. \ref{fig:setup}(b).  A source of entangled photon pairs was built by pumping with a laser two cross-axis BBO crystals. By spontaneous-parametric-down conversion (SPDC), photons are created in the modes $i$ and $s$ in a state $\ket{\psi_0}=\sqrt{p_0}\ket{00}+\sqrt{p_1}\ket{11}$  \cite{Kwiat99}, where the computational basis states $\ket{0}$ and $\ket{1}$ are the horizontal and vertical polarizations of the photons, respectively.  Photons in $i$ are detected with an avalanche-photodiode detector (Det$_i$) after a polarization projection performed by a half wave plate (HWP$_i$) and a polarized beam-splitter   (PBS$_i$). Photons in mode $s$ undergo a unitary process $\mathcal{U}$, implemented by a HWP, and then are detected by Det$_s$ after polarization projection.  Coincident events are counted using a field-programmable-gate-array (FPGA) circuit.	   
  
The reduced initial state of   photons in mode $s$ is a thermal state, which can be written as
\begin{equation}
  \hat{\rho}_0=\frac{1}{1+e^{-\beta}}\ket{0}\bra{0}+\frac{e^{-\beta}}{1+e^{-\beta}}\ket{1}\bra{1},
\end{equation}
where  $\beta=(E_1-E_0)^{-1}\text{ln}(p_0/p_1)$ can be controlled by varying the angle of HWP$_p$.  $E_0$ and $E_1$ are the eigenenergies of the states $\ket{0}$ and $\ket{1}$, respectively.   In what follows, we choose $E_0=0$ and $E_1=1$ without any loss of generality.  It is remarkable that our experimental setup allows  to obtain the probability distribution in Eq. (\ref{7}) by exactly reproducing the  TPM protocol. For this, the photons in $s$  should be measured before and after the unitary $\mathcal{U}$. However, a question that naturally arises is: How do we  carry out the first measurement onto  photon $s$  without destroying it?  To answer it, we utilize the same ideia of Ref. \cite{aguilar},  which consists  in using the initial correlations in the state $\ket{\psi_0}$ and  the post-selection introduced by  detection of coincidences . Thus, by projecting the polarization of $i$ onto the state $\ket{k}_i$,  we guarantee  that the  state of $s$ before $\mathcal{U}$ is $\ket{j}_s$, where $j=k$ both  taking values 0 or 1. Conversely, the second projective measurement onto photon $s$ is performed by HWP$_s$ and PBS$_s$.   It is worth noting that  when the same projective measurement is performed onto $i$ and $s$, the  energy subtraction $E_n-E_m$ is null, meaning a null value for the coherent energy  $\mathcal{C}$.  Similarly, by projecting $i$ onto $\ket{0}$ ($\ket{1}$) and $s$ onto $\ket{1}$ ($\ket{0}$), one gets $\mathcal{C}=1$ ($\mathcal{C}=-1$). 
  
\begin{figure}
    \includegraphics[width=8.5cm]{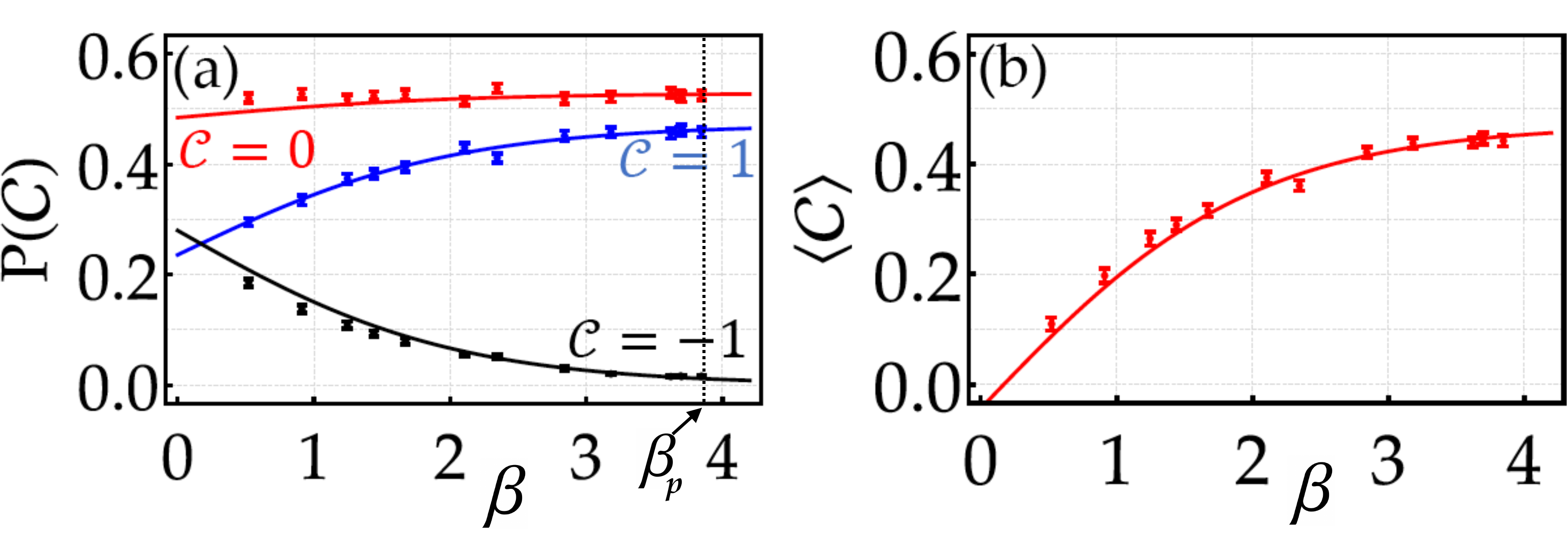}
    \caption{Two-point measurement of the coherent energy. (a) Probability distribution as a function of $\beta$. Experimental and theoretical results for  $P(\mathcal{C}=1)$, $P(\mathcal{C}=-1)$ and $P(\mathcal{C}=0)$ are shown in blue, black and red, respectively. The experimental points at  $\beta_p$ seem to be  in good agreement with the theory. But, the observed small deviations cause large effects on the IFT. (b) Mean coherent energy  $\langle \mathcal{C}\rangle$ as a function of $\beta$, showing that $\langle \mathcal{C}\rangle$  remains non-negative as predicted by Eq. (\ref{6}).}
 \label{Fig:DistribProb}
\end{figure}

%

{\it Results:} In Fig. \ref{Fig:DistribProb}(a) we show $P\left(0\right)$, $P\left(1\right)$ and $P\left(-1\right)$ as a function of $\beta$. Here, one can see that within the error bars the experimental data is  in very good agreement with the theory, plotted by solid  lines. We show that $P\left(0\right)$ (red) takes almost fixed values around 0.5, while $P\left(1\right)$ and $P\left(-1\right)$ (blue and black)  clearly increases and decreases, respectively, as $\beta$ increases. This is physically demonstrating that for decreasing  temperature, there is  higher probability to find the system initially in the fundamental state $\ket{0}$ and then evolving coherently to a higher level $\ket{1}$ than the probability of the opposite transition. In the limit of vanishing temperature, $P\left(-1\right)$ also tends to zero, diminishing the possibility of a downward transition. An alternative way to describe this physical scenario is using the mean coherent energy, which is shown in Fig. \ref{Fig:DistribProb}(b). Here,  an increasing value of $\langle \mathcal{C}\rangle$ means that the transition from a lower to a higher level is more likely than the opposite. In addition, we can also observe here that $\langle \mathcal{C}\rangle$ is always non-negative,  thus indicating the time arrow that dictates the state evolution for this kind of transformation $\mathcal{U}$. The error bars in all plots are obtained by Monte-Carlo simulations of the experimental results obeying Poissonian count statistics. 

\begin{figure}
    \includegraphics[width=8.5cm]{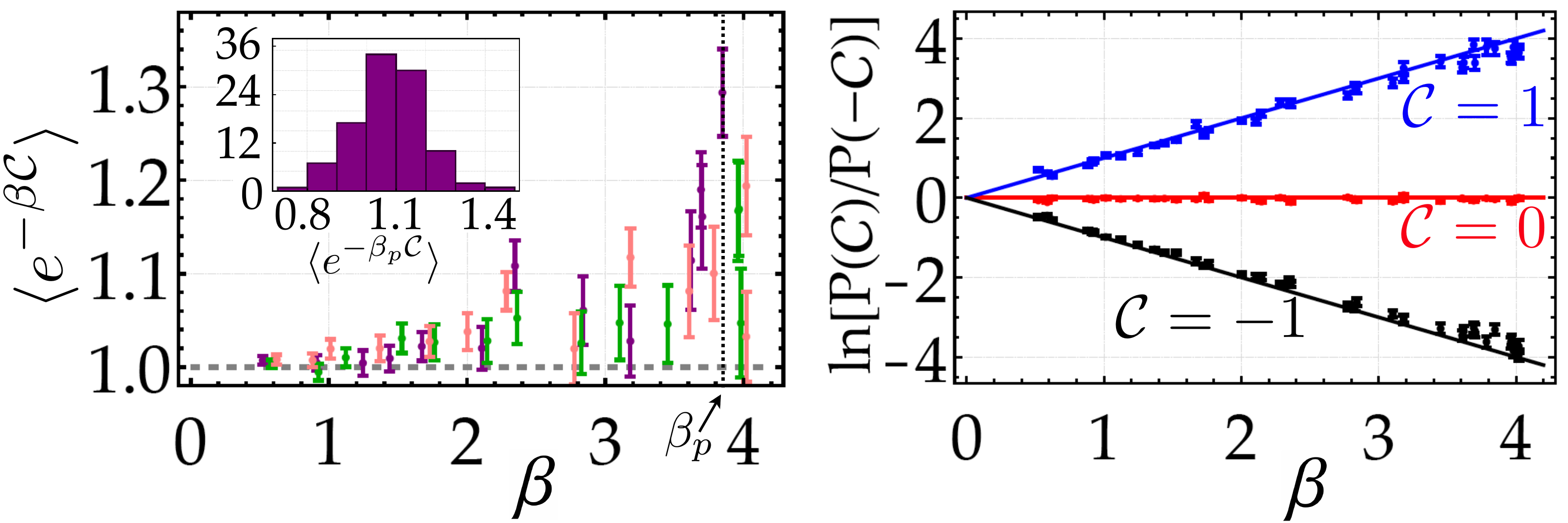}
    \caption{(a) IFT in Eq. (\ref{6}) for three different unitary operations. Experimental data agrees with the predictions from $\beta=0$ up to a value equal to 3. For greater values of $\beta$, measurements present larger deviations, reaching 30$\%$ for $\beta_p$. Inset shows the results  of the IFT at $\beta_p$ as a function of the number N of detected outcomes. The results shown are consistent with the theory. (b) DFT for coherent energy. We plot ln$[P(1)/P(-1)]$ in blue, ln$[P(0)/P(0)]$ in red and ln$[P(-1)/P(1)]$ in black. Excellent linear behavior is observed as expected.    }
    \label{Fig:IntegralFT}
\end{figure}

The results for the IFT in Eq. (\ref{6}) are shown in Fig. \ref{Fig:IntegralFT}(a) for three different unitary operations. It is observed that for  $\beta$ close to 0 (very high temperatures), the experimental results are in very good agreement with the theoretical predictions (dashed line). However,  large deviations are obtained as $\beta$ increases. To investigate the origin of this, in the Supplemental Material we calculate the explicit dependence of  $\langle e^{-\beta \mathcal{C}}\rangle$ on $\beta$ \cite{Supplementar}. We show that one of the possible realizations  contributing to the IFT is proportional to  $e^{\beta}$, which diverges for large $\beta$. To circumvent this issue, the theoretical predictions in Fig. \ref{Fig:DistribProb}(a) shows that this realization (black curve) tends to be unlikely when $\beta$ goes to infinity. Nevertheless, in an experimental context, we could have noisy detections due to imperfections of our apparatus, which could bring undesirable effects in our results   even for  $\beta$ less than five. For instance,  it can be  observed in Fig. \ref{Fig:DistribProb}(a) that the experimental point of $P(-1)$ at $\beta_p$ is slightly above the theoretical curve, generating deviations of up to 30$\%$ in the IFT with respect to the expected unity. To better understand the nature of the deviations  at $\beta_p$, we repeat our measurements more than one hundred times. The results are observed in the inset of Fig. \ref{Fig:IntegralFT}(a). From the histogram one can see that most of the results of the measurements of the IFT fall in the interval between  1 and 1.2, obtaining a mean value of 1.07 and a standard deviation of 0.12. This shows that for $\beta=\beta_p$ the results are actually compatible with the expectations within the error bars.   

The experimental results for the coherent energy DFT are shown in Fig. \ref{Fig:IntegralFT}(b). For this case, it was also necessary to obtain the probabilities for the time-reversed process, which for polarization measurements corresponds to rotate the HWP optical axis by 90 degrees with respect to the original (forward) process. We  plot  $\text{ln}[P(1)/P(-1)]$, $\text{ln}[P(0)/P(0)]$ and $\text{ln}[P(-1)/P(1)]$ in blue, red and black, respectively. As expected, the plots in the logarithmic scale present linear behavior.   Deviations of the data with respect to the predictions for larger values of $\beta$ when $\mathcal{C}\neq 0$ are also observed. This effect is analogous to that analyzed previously: small deviations in $P(-1)$ bring large deviations in the FT.


{\it Discussion:} Throughout this work, we study the coherent energy from both the fundamental and experimental side. We theoretically deduced two different QFT for the coherent energy, which indicates that the average internal energy can only increase for a system in thermal equilibrium undergoing an arbitrary unitary evolution that preserves the energy spectrum. In the present scenario, the two possible types of energy transfer occurring in a closed system are work and coherent energy. Nevertheless, from Eq. (1) it is clear that the average work can be both positive or negative, depending only on the increase or decrease of the energy eigenvalues due to the quantum dynamics. This leads us to conclude that the unidirectional feature  revealed by the inequality $\langle \mathcal{C} \rangle \geq 0$ gives to the coherent energy the significance of  an ``arrow of time''  for unitary evolutions involving thermal states, i.e., when the entropy production is not present. This is not a trivial result, specially for arbitrary Hilbert space dimensions.    We performed an experiment, which allows the implementation of the TPM protocol, to investigate the obtained QFT for different values of temperature. Our  experimental result promote the relevance of using initially correlated systems to perform the first measurement of the TPM protocol, such as it was explored recently in Ref. \cite{aguilar}. The experimental data agree well with the theory for a large temperature range. To the best of our knowledge, these results are the first experimental verification of a QFT for a quantity directly linked with the coherence of the quantum state.  This opens up a new avenue for experimental research where the coherences of the states are considered in far from equilibrium dynamics.

\begin{acknowledgements}
The authors acknowledge financial support from the Brazilian agencies CNPq (PQ Grants No. 307058/2017-4, No. 303451/2019-0   and and INCT-IQ 246569/2014-0).
GHA  acknowledges FAPERJ (Grant No. 210.069/2020). BLB acknowledges Pronex/Fapesq-PB/CNPq, Grant No. 0016/2019, and PROPESQ/PRPG/UFPB (PIA13177-2020).

\end{acknowledgements}


%
%
   

%

\pagebreak
\widetext

\newpage 
\begin{center}
\vskip0.5cm
{\Large Supplemental Material}
\end{center}
\vskip0.4cm

\setcounter{section}{0}
\setcounter{equation}{0}
\setcounter{figure}{0}
\setcounter{table}{0}
\setcounter{page}{1}
\renewcommand{\theequation}{S\arabic{equation}}
\renewcommand{\thefigure}{S\arabic{figure}}

\vskip0.7cm

In this supplemental material we provide details of the fluctuation theorems (FT), derived in the main text, for the special case of two-dimensional systems such as the polarization of photons.  In particular, the present results shed some light on the interplay between FT and the study of optical quantum information processing. 

\section{Quantum optical perspective of the Coherent Energy Fluctuation Theorems}

\subsection{Integral Fluctuation theorem}

We begin with a quantum description of the results obtained in our experiment for the data related with the integral fluctuation theorem (IFT). As discussed in the main text, the photons in mode $s$ are initially prepared in the (reduced) initial state

\begin{equation}
\label{S1}
\rho_{0} = p_{0} \ket{0}\bra{0} + p_{1} \ket{1}\bra{1} =\frac{1}{1+ e^{- \beta}}\ket{0}\bra{0} + \frac{e^{- \beta}}{1+ e^{- \beta}}\ket{1}\bra{1},
\end{equation}
where $\beta$ represents the inverse temperature, and the computational basis $\ket{0}$ and $\ket{1}$
are the horizontal and vertical polarization states,
respectively. In realizing the two-point measurement (TPM) protocol, we have seen that the coherent energy $\mathcal{C}$, which is the difference between the second and first outcomes of the energy measurements, can assume three different values: $\mathcal{C}_{1} = 0$, $\mathcal{C}_{2} = -1$ and $\mathcal{C}_{3} = 1$. In this case, we have that  

\begin{equation}
\label{S2}
\langle e^{- \beta \mathcal{C} }\rangle = \sum_{i = 1}^{3} e^{- \beta \mathcal{C}_{i}} P(\mathcal{C}_{i}),
\end{equation}
where $P(\mathcal{C}_{i})$ is the probability distribution for $\mathcal{C}$.

From the quantum mechanical probabilistic rules, it is easy to see that
\begin{equation}
\label{S3}
P(0) = |\braket{0|\hat{\mathcal{U}}|0}|^{2}p_{0} + |\braket{1|\hat{\mathcal{U}}|1}|^{2}p_{1},
\end{equation}
\begin{equation}
\label{S4}
P(-1) = |\braket{0|\hat{\mathcal{U}}|1}|^{2}p_{1},
\end{equation}
\begin{equation}
\label{S5}
P(1) = |\braket{1|\hat{\mathcal{U}}|0}|^{2}p_{0},
\end{equation}
with $p_{0}$ and $p_{1}$ given as in Eq.~(\ref{S1}), and $\hat{\mathcal{U}}$ an arbitrary unitary operation that keeps the energy levels of the system fixed. By substitution of Eqs.~(\ref{S3}) to (\ref{S5}) into Eq.~(\ref{S2}), we obtain that
\begin{align}
\label{S6}
\langle e^{- \beta \mathcal{C} }\rangle =& \frac{1}{1+ e^{- \beta}} \left[ |\braket{0|\hat{\mathcal{U}}|0}|^{2} +|\braket{0|\hat{\mathcal{U}}|1}|^{2} + e^{-\beta} (|\braket{1|\hat{\mathcal{U}}|0}|^{2} + | \braket{1|\hat{\mathcal{U}}|1}|^{2}) \right],
\end{align}
which can be rewritten as $\langle e^{- \beta \mathcal{C} }\rangle =P(0)+P(1)e^{-\beta}+P(-1)e^{\beta}$. Since $\hat{\mathcal{U}}$ is unitary, the rows of its matrix representation form an orthonormal basis of $\mathbb{C}^2$ \cite{nielsen}, i.e., 
\begin{equation}
\label{S7}
 |\braket{0|\hat{\mathcal{U}}|0}|^{2} +|\braket{0|\hat{\mathcal{U}}|1}|^{2} =|\braket{1|\hat{\mathcal{U}}|0}|^{2} + | \braket{1|\hat{\mathcal{U}}|1}|^{2} = 1.
\end{equation}
By using this result in Eq.~(\ref{S6}), we get 
\begin{equation}
\label{S8}
\langle e^{- \beta \mathcal{C} }\rangle = 1.
\end{equation}
This confirms the validity of the IFT for two-level systems.  

\subsection{Detailed Fluctuation theorem}

Now we turn to the study of the detailed fluctuation theorem (DFT) derived in the main text, for the case of two-level systems. Again, from the quantum mechanical probabilistic rules, we have that

\begin{equation}
\label{S9}
\frac{P(\mathcal{C}_{1})}{\Tilde{P}(\mathcal{-C}_{1})} = \frac{P_{0}|\braket{0|\hat{\mathcal{U}}|0}|^{2} + P_{1}|\braket{1|\hat{\mathcal{U}}|1}|^{2}}{P_{0}|\braket{0|\hat{\mathcal{U}^{-1}}|0}|^{2} + P_{1}|\braket{1|\hat{\mathcal{U}^{-1}}|1}|^{2}} = 1,
\end{equation}
where we used the fact that unitary operators satisfy $\mathcal{U}^{-1} = \mathcal{U}^{\dagger}$. Similarly, we have that
\begin{equation}
\label{S10}
\frac{P(\mathcal{C}_{2})}{\Tilde{P}(\mathcal{-C}_{2})} = \frac{P_{1}|\braket{0|\hat{\mathcal{U}}|1}|^{2}}{P_{0}|\braket{1|\hat{\mathcal{U}^{-1}}|0}|^{2}} = e^{- \beta},
\end{equation}
and
\begin{equation}
\label{S10}
\frac{P(\mathcal{C}_{3})}{\Tilde{P}(\mathcal{-C}_{3})} = \frac{P_{0}|\braket{1|\hat{\mathcal{U}}|0}|^{2}}{P_{1}|\braket{0|\hat{\mathcal{U}^{-1}}|1}|^{2}} = e^{\beta}.
\end{equation}
The results of Eqs.~(\ref{S9}) to (\ref{S10}) can be summarized as follows:  
\begin{align}
\label{S12}
\frac{P(\mathcal{C}_{i})}{\Tilde{P}(-\mathcal{C}_{i})} = e^{\beta \mathcal{C}_{i}}.
\end{align}
This proves theoretically that the data of our experiment should also satisfy the 
DFT derived in the main text.

\subsection{Arrow of time relation}

The unidirectional aspect of the coherent energy was clearly demonstrated through the relation $\langle \mathcal{C} \rangle \geq 0$, which could be easily derived from the IFT in the main text. Still, much physical insight into this ``quantum arrow of time'' is provided if we investigate this relation directly. Such an investigation is not a trivial task if we deal with quantum systems of arbitrary Hilbert space dimension. However, we address this issue here by using the simple qubit case studied in our experiment. Since our system is restricted to quantum dynamics that are generated by unitary operations which preserve the structure of the energy levels of the system, all change in the internal energy is due to coherent energy transfer. Thus, we can write $\langle \mathcal{C} \rangle = \langle \hat{H} \rangle_{f} - \langle \hat{H} \rangle_{0}$, where $\langle \hat{H} \rangle_{0}$ and $\langle \hat{H} \rangle_{f}$ are the averages of the Hamiltonian at the beginning and the end of the process, respectively. This is given by 
\begin{align}
\label{S13}
\langle \mathcal{C} \rangle = tr[\hat{\rho}_{f} \hat{H}] - tr[\hat{\rho}_{0} \hat{H}] = tr[\hat{\mathcal{U}} \hat{\rho}_{0} \hat{\mathcal{U}}^{\dagger} \hat{H}] - tr[\hat{\rho}_{0} \hat{H}],
\end{align}
with $\hat{\rho}_{0}$ and $\hat{\rho}_{f}$ being the initial and final states, respectively, and $\hat{\mathcal{U}}$ the unitary time evolution operator. In the case of our experiment, we have $\hat{\rho}_{0}$ as given in Eq.~(\ref{S1}) and $\hat{H} = \ket{1}\bra{1}$.

The arbitrary unitary transformation $\hat{\mathcal{U}}$ that keeps the energy levels unchanged is here characterized by a general state rotation in the Bloch sphere, which can be represented by a rotation around the $y$-axis by an angle $\theta$, followed by a rotation around the $z$-axis by an angle $\phi$, with $0 \leq \theta \leq \pi$ and $0 \leq \phi \leq 2\pi$ \cite{sakurai}. Then, we can use a decomposition given by $\hat{\mathcal{U}}(\theta,\phi)= \exp(-i \hat{\sigma}_{z} \phi/2) \exp(-i \hat{\sigma}_{y} \theta/2)$, where $\hat{\sigma}_{y} = -i \ket{0}\bra{1} +i\ket{1}\bra{0}$ and $\hat{\sigma}_{z} = \ket{0}\bra{0} - \ket{1}\bra{1}$ are Pauli operators. Therefore, after some algebra we find that
\begin{align}
\label{S14}
\hat{\mathcal{U}}(\theta,\phi)= \cos \left( \frac{\theta}{2}  \right) \left[e^{-i \phi/2} \ket{0}\bra{0} + e^{i \phi/2} \ket{1}\bra{1}\right] + \sin \left( \frac{\theta}{2}  \right) \left[-e^{-i \phi/2} \ket{0}\bra{1} + e^{i \phi/2} \ket{1}\bra{0}\right].
\end{align}
These results allow us to write
\begin{align}
\label{S15}
 tr[\hat{\mathcal{U}} \hat{\rho}_{0} \hat{\mathcal{U}}^{\dagger} \hat{H}] = p_{0} \sin^2 \left( \frac{\theta}{2}  \right) + p_{1} \cos^2 \left( \frac{\theta}{2}  \right)
\end{align}
and
\begin{align}
\label{S16}
tr[\hat{\rho}_{0} \hat{H}] = p_{1}.
\end{align}

By substitution of Eqs.~(\ref{S15}) and~(\ref{S16}) into Eq.~(\ref{S13}) we obtain that
\begin{align}
\label{S17}
\langle \mathcal{C} \rangle = \sin^2 \left( \frac{\theta}{2}  \right) \left[p_{0} - p_{1} \right].
\end{align}
Finally, if we express $p_{0}$ and $p_{1}$ in terms of the inverse temperature $\beta$, according to Eq.~(\ref{S1}), this result is reduced to the following expression:  
\begin{align}
\label{S18}
\langle \mathcal{C} \rangle = \sin^2 \left( \frac{\theta}{2}  \right) \tanh \left( \frac{\beta}{2}  \right).
\end{align}
Note that since $\beta \geq 0$, this relation proves that our experimental data should indeed satisfy the relation $\langle \mathcal{C} \rangle \geq 0$, which is our indicator of the arrow of time. In fact, our experimental data for $\langle \mathcal{C} \rangle $ as a function of $\beta$ shown in Fig. 2(b) of the main text fits very well with the result of Eq.~(\ref{S18}) for $\theta \approx 86.6^{\circ}$. From Eq.~(\ref{S18}), we observe that two limit cases can be considered: i) $\langle \mathcal{C} \rangle = 0$, which occurs either in the absence of quantum processes ($\theta = 0$, which according to Eq.~(\ref{S14}) implies $\hat{\mathcal{U}}(\theta,\phi) = \hat{\mathbb{I}}$, up to an irrelevant global phase factor), or in the limit of infinite temperature ($\beta \rightarrow 0$); and ii) $\langle \mathcal{C} \rangle = 1$, which occurs in the limit of vanishing temperature ($\beta \rightarrow \infty$), and the unitary operation is such that $\theta = \pi$. 

\begin{figure}[ht]
\centerline{\includegraphics[width=6.5cm]{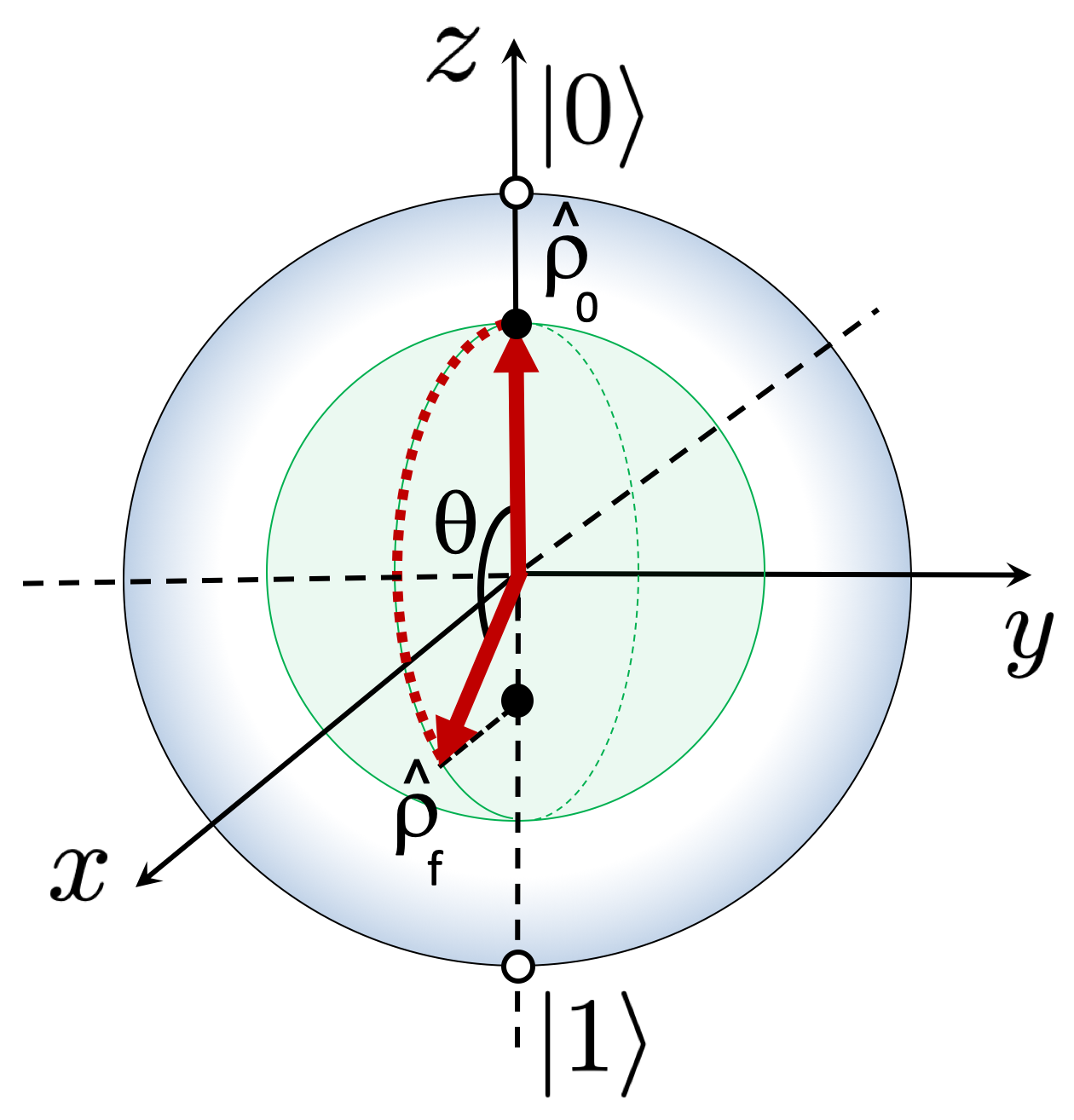}}
\caption{Bloch sphere description of a rotation operation about the $y$-axis by an angle $\theta$ acting on a thermal state (upward vector), which can be implemented by a half-wave plate (HWP). In this perspective, the coherent energy transferred in the process is represented by the difference between the initial and final $z$ components of the Bloch vector. This can be visualized as the distance between the filled circles. Open circles represent the basis states $\ket{0}$ and $\ket{1}$.}
\label{setup}
\end{figure}

In our experiment, the rotation of the Bloch vector about the $y$-axis by an angle $\theta$ is implemented by a HWP with the fast axis oriented at  $\theta/4$ with respect to the horizontal, whereas rotations about the $z$-axis can be implemented with an arrangement of quarter- and half-wave plates \cite{kok}. As can also be seen in Eq.~(\ref{S18}), $\langle \mathcal{C} \rangle$ only depends on the rotation angle $\theta$. Some insight can also be gained about the nature of this quantum dynamics by using the Bloch sphere representation. In this framework, the Bloch vector $\Vec{a} = (a_{x},a_{y},a_{z})$ uniquely represents a general (mixed) qubit state as given by
\begin{align}
\label{S19}
\hat{\rho}  = 
\begin{pmatrix}
    1 + a_{z} & a_{x} - i a_{y} \\
    a_{x} + i a_{y} & 1 - a_{z}\\
\end{pmatrix},
\end{align} 
such that the internal average energy of the state becomes 
\begin{align}
\label{S120}
\langle \hat{H} \rangle = tr[\hat{H} \hat{\rho}] = 1 - a_{z}.
\end{align}
With this, the average coherent energy becomes 
\begin{align}
\label{S21}
\langle \mathcal{C} \rangle = tr[\hat{\rho}_{f} \hat{H}] - tr[\hat{\rho}_{0} \hat{H}] =  a^{(0)}_{z}-a^{(f)}_{z}  = -\Delta a_{z},
\end{align}
which is minus the change in the $z$ component of the Bloch vector due to the rotation operation. For instance, in the case of the initial state in Eq.~(\ref{S1}), $a^{(0)}_z$ results in $(p_0-p_1)/2$, while for the evolved state   $\hat{\mathcal{U}} \hat{\rho}_{0} \hat{\mathcal{U}}^{\dagger}$ we obtain  $a^{(f)}_{z}=a^{(0)}_z\cos{\theta}\leq a^{(0)}_z$. This means that any unitary evolution diminishes the $z$ component of the Bloch vector, thus indicating a direction in which time evolves. In other words, since    the length of the vector is fixed in the unitary processes, the rotated state will always have the $z$ component smaller than that of $\hat{\rho}_{0}$,  represented by the Bloch vector pointing upwards in Fig. S1. These results are all equivalent with  $\langle \mathcal{C} \rangle \geq 0$, establishing a clear relation between the arrow of time and the mean coherent energy for unitary processes. It is also worth mentioning that these results can be extended for systems with larger Hilbert spaces, but without the pictorial representation of the Bloch sphere.

\end{document}